\UseRawInputEncoding
\documentclass[
aps,
prb,
twocolumn,
superscriptaddress,
nofootinbib,
amsmath,amssymb,
floatfix]{revtex4-2}

\usepackage{graphicx}%
\usepackage{dcolumn}%
\usepackage{bm}%
\usepackage[%
    colorlinks=true,
    pdfborder={0 0 0},
    linkcolor= black,
    citecolor= black
]{hyperref}
\usepackage{etoolbox}
\usepackage[flushleft]{threeparttable}
\newcommand{\convert}[1]{$|e0\rangle\leftrightarrow|g1\rangle$}
\makeatletter

\makeatletter

\makeatletter
\def\@email#1#2{%
 \endgroup
 \patchcmd{\titleblock@produce}
  {\frontmatter@RRAPformat}
  {\frontmatter@RRAPformat{\produce@RRAP{*#1\href{mailto:#2}{#2}}}\frontmatter@RRAPformat}
  {}{}
}%
\makeatother
\begin{document}

    \title{Parametrically-controlled microwave-photonic interface for the fluxonium}

    \author{Ke Nie}
    \affiliation{%
    Department of Physics, University of Illinois Urbana-Champaign, Urbana, IL 61801}
    \author{Aayam Bista}
    \affiliation{%
    Department of Physics, University of Illinois Urbana-Champaign, Urbana, IL 61801}
    \author{Kaicheung Chow}
    \affiliation{Holonyak Micro $\&$ Nanotechnology Lab, University of Illinois Urbana-Champaign, Urbana, IL 61801}
    \author{Wolfgang Pfaff}
    \affiliation{%
     Department of Physics, University of Illinois Urbana-Champaign, Urbana, IL 61801}
     \affiliation{Materials Research Laboratory,
University of Illinois at Urbana-Champaign, Urbana, IL 61801}
    \author{Angela Kou} 
    \affiliation{%
    Department of Physics, University of Illinois Urbana-Champaign, Urbana, IL 61801}
    \affiliation{Holonyak Micro $\&$ Nanotechnology Lab, University of Illinois Urbana-Champaign, Urbana, IL 61801}
    \affiliation{Materials Research Laboratory,
    University of Illinois at Urbana-Champaign, Urbana, IL 61801}

    \date{\today}

    \begin{abstract}
        Converting quantum information from stationary qubits to traveling photons enables both fast qubit initialization and efficient generation of flying qubits for redistribution of quantum information. This conversion can be performed using cavity sideband transitions. In the fluxonium, however, direct cavity sideband transitions are forbidden due to parity symmetry. Here we circumvent this parity selection rule by using a three-wave mixing element to couple the fluxonium to a resonator. We experimentally demonstrate a scheme for interfacing the fluxonium with traveling photons through microwave-induced parametric conversion. We perform fast reset on the fluxonium qubit, initializing it with $>95\%$ ground state population. We then implement controlled release and temporal shaping of a flying photon, useful for quantum state transfer and remote entanglement. The simplicity and flexibility of our demonstrated scheme enables fluxonium-based remote entanglement architectures.

    \end{abstract}

    \maketitle

    \section{Introduction}
    Interfacing stationary logical qubits and flying photons can serve multiple purposes \cite{DiVincenzo_2000}, such as implementing on-demand transfer and remote entanglement of quantum bits in a modular quantum system \cite{PhysRevLett.78.3221, Korotkov_2011}, fast evacuation of qubit excitations for initialization \cite{2017NatPh..13..882P}, and efficient generation of non-classical single-photon states \cite{ Keller2004ContinuousGO, Barros_2009, Stute_2013, PRXQuantum.2.020331, Houck_2007, Kuhn_2002, doi:10.1126/science.1095232, M_cke_2013}. 
    In circuit quantum electrodynamics (cQED) \cite{Blais_2021}, high-fidelity qubit-photonic
    interfaces have been built due to the availability of strong coupling between superconducting circuits and microwave photons. Experiments interfacing a transmon qubit \cite{Koch_2007} with traveling photons have demonstrated fast unconditional qubit reset \cite{Magnard_2018, Zhou_2021} and deterministic generation and temporal shaping of single photons for remote entanglement \cite{ PhysRevX.4.041010, Campagne_Ibarcq_2018, Kurpiers_2018}.             

    In recent years, the fluxonium \cite{Manucharyan_2009} has emerged as a promising superconducting qubit due to its remarkably long coherence times \cite{PhysRevX.9.041041, PhysRevLett.130.267001} and high gate fidelities \cite{PhysRevX.13.031035, Moskalenko_2022}. A coherent fluxonium-photonic interface would be beneficial for fluxonium quantum processing on multiple fronts. First, efficient initialization of the fluxonium is important as the low transition frequency of the fluxonium results in a highly-mixed thermal population. Second, faithful mapping between a fluxonium excitation and a flying qubit would allow the realization of modular quantum systems linked via traveling microwave photons using the fluxonium. A typical method for engineering this interface involves converting a fluxonium excitation into a fast-decaying resonator photon through pumping of cavity sideband transitions \cite{wang2024efficient, Zhang_2021, Vool_2018}.
    However, direct sideband transitions between fluxonium-resonator states of the same total parity are forbidden under a linear microwave drive. 
    Recent experiments have reported multiple ways of circumventing the parity selection rule forbidding these direct transitions by flux-biasing the fluxonium qubit away from half flux \cite{wang2024efficient}, using microwave-driven multi-photon transitions \cite{Zhang_2021}, and coupling the fluxonium to a resonator via a third-order mixing element known as the Superconducting Nonlinear Asymmetric Inductive eLement (SNAIL) \cite{Frattini_2017, Vool_2018}. The above-mentioned experiments have been solely focused on fluxonium initialization and cooling. 

    \par
    In this Letter, we report an experiment interfacing the fluxonium qubit with traveling photons using parametric conversion in a fluxonium-resonator system that enables both fluxonium initialization and efficient generation of flying qubits. We use a SNAIL-based resonator as an intermediary coupling element to coherently convert between fluxonium excitations and microwave photons. We control the coupling rate by modulating the pump amplitude, which allows for fast switching between different quantum information tasks. In particular, we perform fast reset of the fluxonium qubit in the strong coupling regime. We then operate in the weak coupling regime to controllably release and temporally shape single traveling photons. Our scheme provides a straightforward pathway for both fluxonium initialization and for incorporating the fluxonium qubit into remote entanglement schemes based on traveling microwave photons. 
            
        \begin{figure}
        \centering
            \includegraphics{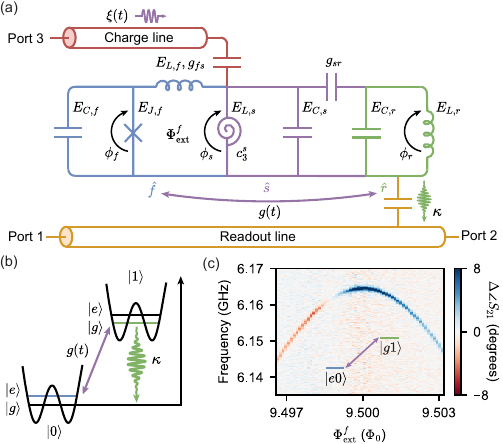}
            \caption{\label{fig:1} 
                (a) Circuit schematic of the device. A fluxonium (blue) is inductively coupled to a SNAIL resonator (purple), which couples capacitively to a readout resonator (green). The SNAIL's three-wave mixing term mediates the conversion of a fluxonium excitation to a resonator photon at coupling rate $g(t)$. The readout resonator emits photons to a coupled transmission line (orange) with decay rate $\kappa$. Pump tones for qubit control and parametric conversion are applied through a separate charge line (red). (b) Energy level structure of the fluxonium-resonator system shown in (a) used for interfacing a fluxonium excitation with a microwave photon.    
                (c) Two-tone spectroscopy data of the $|e0\rangle\leftrightarrow|g1\rangle$ transition near $\Phi_\text{ext}^f/\Phi_0 = 9.5$. The change in the sign of the resonator phase response is due to a flip in the direction of the dispersive shift.}
        \end{figure}

        \section{Fluxonium-SNAIL-resonator system}\label{Fluxonium-SNAIL-resonator system}

        \par
        The main challenge faced when building a coherent fluxonium-photonic interface is achieving strong, tunable coupling in a fluxonium-resonator system. Direct conversion between the fluxonium and resonator photons is forbidden due to the parity selection rule in linearly-coupled fluxonium-resonator systems. Strong coupling is made further difficult by the large detuning between the fluxonium computational states and the resonator states. These barriers can be overcome by engineering third-order nonlinear coupling between the fluxonium and a resonator via a SNAIL (first demonstrated in Ref. \cite{Vool_2018}). The SNAIL's third-order nonlinearity breaks the parity symmetry of the wavefunctions in the composite Hilbert space. Furthermore, operating the SNAIL at a similar transition frequency to the resonator while inductively coupling it to the fluxonium through the $\phi$ operator effectively allows the SNAIL to bridge the detuning gap between the fluxonium and the resonator. The SNAIL is then capable of strong coupling to both the fluxonium computational states and the resonator simultaneously, which makes it an excellent mediator for a fluxonium-photonic interface. 
        
        Our circuit for implementing third-order nonlinear coupling of a fluxonium-resonator
        system is shown in Fig. \ref{fig:1}(a). We build a system of three coupled modes: the fluxonium qubit mode ($\hat{f}$), an intermediary SNAIL mode ($\hat{s}$), and a readout resonator mode ($\hat{r}$) coupled to a transmission line with loss rate $\kappa/2\pi=0.4$ MHz. The system Hamiltonian up to third order is given by 
        \begin{multline}\label{Hamiltonian}
            \hat{H}_\text{sys}/\hbar= \sum_{i=f,s,r} (4E_{C,i}\hat{n}_i^2+\frac{1}{2}E_{L,i}\hat{\phi}_i^2) \\-E_{J,f}\cos{(\hat{\phi}_f-\phi_\text{ext}^f)}   +c_3^s\hat{\phi}_s^3+g_{fs}\hat{\phi}_f\hat{\phi}_s+g_{sr}\hat{n}_s\hat{n}_r.
        \end{multline}      
        Here $E_{C,i}, E_{L,i}, E_{J,i}$ are the charging, inductive, and Josephson energies, respectively (see Appendix \ref{A1} for system parameters and time domain measurements). The Cooper pair number and phase operator are denoted by $\hat{n}_i$ and $\hat{\phi_i}$. The subscripts $f, s, r$ represent the fluxonium qubit, the SNAIL resonator, and the readout resonator. We also introduce 
        $\phi_\text{ext}^f  \equiv 2\pi\Phi_\text{ext}^f/\Phi_0$, where $\Phi_\text{ext}^f$ is the external flux through the fluxonium qubit. The coupling strength between each neighboring element is captured by $g_{ij}$ in the Hamiltonian, and $c_3^s$ represents the third-order coefficient of the SNAIL. Importantly, the SNAIL's third-order term enables all energy-conserving three-wave mixing processes. We can then apply a microwave pump with frequency matching the detuning between the fluxonium mode and the readout resonator mode to realize the effective system Hamiltonian,
        \begin{equation}
            \hat{H}_\mathrm{conv}/\hbar=g(t)\hat{f}\hat{r}^\dag+h.c..
        \end{equation}
        Here the coupling rate $g(t)$ is given by 
        \begin{equation}\label{rabi}
            g(t)=6c_3^s\xi(t)\phi_{f,\mathrm{zpf}}\phi_{s,\mathrm{zpf}}\phi_{r,\mathrm{zpf}},
        \end{equation}   
        with $|\xi(t)|^2$ being the number of pump photons in the SNAIL resonator, and $\phi_{i,\mathrm{zpf}}$ denoting the zero-point fluctuations of the flux associated with each mode (Appendix \ref{A3}).
        This Hamiltonian allows us to drive transitions of the form $|i, n\rangle\leftrightarrow|j,n+1\rangle, ~i\neq j$ and $n>0$, where $i,j$ denote any neighbouring eigenstates of the fluxonium with $|i\rangle$ being the higher eigenstate, and $n$ denotes the number of photons in the readout resonator. 
        By directly driving the $|e0\rangle\leftrightarrow|g1\rangle$ transition, excitations of the fluxonium can be parametrically converted into a photon in the readout resonator, which is then emitted at the decay rate $\kappa$ (Fig. \ref{fig:1}(b)). 

        \par
        The rate of conversion is controlled by the pump amplitude $\xi(t)$ and the third-order coefficient $c_3^s$. A strong microwave drive on the SNAIL induces a significant ac-Stark shift on the transition (see Appendix \ref{A3} for further details on the system Hamiltonian and ac-Stark shift), which complicates pulse calibration due to the Stark shift changing non-negligibly during the pulse rise time. We therefore choose to work with a large $c_3^s$, which can be obtained by operating at nonzero $\Phi_\text{ext}^s$ through the SNAIL. We flux bias the system to $\Phi_\text{ext}^f/\Phi_\text{0}=9.5$, corresponding to $\Phi_\text{ext}^s/\Phi_\text{0}=0.27$ (Appendix \ref{A1}). Spectroscopy of the desired $|e0\rangle\leftrightarrow|g1\rangle$  
        transition at this flux point is shown in Fig.~\ref{fig:1}(c) (see Appendix \ref{A2} for wiring). 
        We perform dispersive readout on the fluxonium enabled by the last two, always-on coupling terms in Eq. (\ref{Hamiltonian}). 
        We note the clear visibility of this transition at the half-flux sweet spot of the fluxonium, indicating that the parity selection rule has been broken \cite{Vool_2018}. The expected $c^s_3/2\pi$ is simulated to be 350~MHz at this point.

        \begin{figure}[t]
        \centering
            \includegraphics{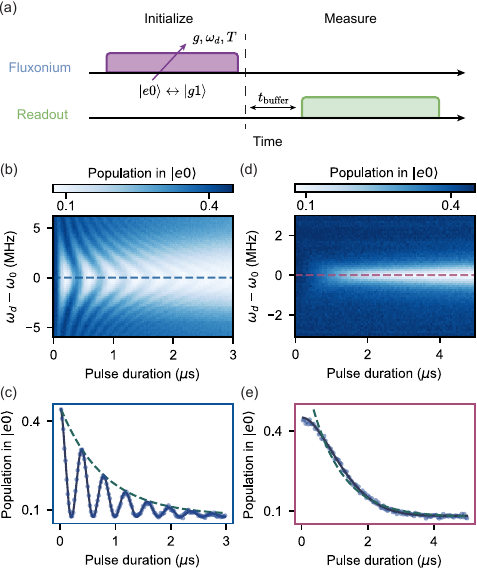}%
            \caption{\label{fig:2} 
                (a) The pulse sequence for characterizing coherence and tunability of the $|e0\rangle\rightarrow|g1\rangle$ transition. A square pulse (purple) with frequency $\omega_d$ is applied to the system to drive the $|e0\rangle\rightarrow|g1\rangle$ transition. A wait time of $t_\text{buffer}$ is included to ensure that the photon leaves the resonator before readout (green).  
                (b) Rabi oscillations of the $|e0\rangle\leftrightarrow|g1\rangle$ transition under a strong pump tone ($g\gg\kappa$) as a function of detuning. (c) Measured qubit $|e0\rangle$ population vs. pulse duration data from (b) at zero detuning. The data is fit using the master equation (solid). The effective qubit decay rate is given by $\kappa/2$ in the strong coupling regime (dashed).
                (d) The time evolution of the $|e0\rangle$ population under a weak pump tone ($g\ll\kappa$) as a function of detuning between the pump and the transition frequency $\omega_d - \omega_0$. (e) Measured qubit $|e0\rangle$ population vs. pulse duration data from (d) at zero detuning. A simple exponential decay with $\Gamma_\mathrm{eff}=4g^2/\kappa$ is plotted for reference (dashed).
            }
        \end{figure}

    \section{Coherent Driving}
        \par
        Here we demonstrate that our parametric scheme can reach different coupling regimes useful for fast reset and generating flying photons. %
        We characterize the transition $|e0\rangle\leftrightarrow|g1\rangle$ by driving it with a square pulse of varying duration and power (Fig. \ref{fig:2}). We measure the qubit population in time as a function of detuning $\omega_d-\omega_0$, where $\omega_d, \omega_0$ represent the driving frequency and the transition frequency, respectively (Fig. \ref{fig:2}(a)). As the system evolves coherently as a superposition of $|e0\rangle$ and $|g1\rangle$, $|g1\rangle$ decays back to $|g0\rangle$ and emits the photon field into the transmission line at an effective decay rate depending on the relative strength of $g$ and $\kappa$.
        We then wait for time $t_\text{buffer}=5$~$\mu$s ($\sim10\kappa^{-1}$), to ensure that the photon has left the resonator to avoid changes to the qubit readout as well as frequency shifts based on the cavity state before any subsequent operation on the fluxonium. 

        We first characterize the operation of driving the $|e0\rangle\leftrightarrow|g1\rangle$ transition in the strong coupling regime ($g\gg\kappa$), which is useful for fast qubit reset. Under a strong pump tone, the system undergoes fast Rabi oscillations between $|e0\rangle$ and $|g1\rangle$ as shown in Fig. \ref{fig:2}(b). We fit the data at zero detuning (Fig. \ref{fig:2}(c)) using the system master equation (Appendix C). From this fit, we find $g/2\pi=1.27 \text{ MHz} \gg \kappa/2\pi$ with exponential decay rate $\Gamma_\mathrm{eff}=\kappa/2$ given that the qubit lifetime $T_1\gg1/\kappa$, which indicates that the system is in the strong coupling regime.

        \par
        We then characterize the $|e0\rangle\leftrightarrow|g1\rangle$ transition in the weak coupling regime, which is useful for efficient generation and temporal shaping of flying photons. Under a relatively weak pump, the state $|e0\rangle$ is slowly converted into $|g1\rangle$ with linewidth $\kappa$. In this regime, we can tune the effective decay rate $\Gamma_\mathrm{eff} = 4g^2/\kappa$ via the strength of our pump, which controls $g$. 
        In Fig. \ref{fig:2}(d), we show the qubit $|e0\rangle$ population evolution in time as a function of detuning $\omega_d-\omega_0$ when the $|e0\rangle\leftrightarrow|g1\rangle$ transition is driven by a relatively weak pump tone. The data at zero detuning is plotted separately in Fig. \ref{fig:2}(e).
        A fit to the master equation (Appendix \ref{A3}) yields a coupling rate $g/2\pi=0.13\text{ MHz} \ll \kappa/2\pi$. The qubit decay exhibits no oscillations and resembles that of standard Purcell loss after the ring-down time $1/\kappa$ with effective decay rate $\Gamma_\mathrm{eff}=4g^2/\kappa$. Now with a well-characterized ability to drive and tune the $|e0\rangle\leftrightarrow|g1\rangle$ transition, we can use this transition for fast reset and controlled release of single photons.

        \begin{figure}[t]
        \centering
        \includegraphics{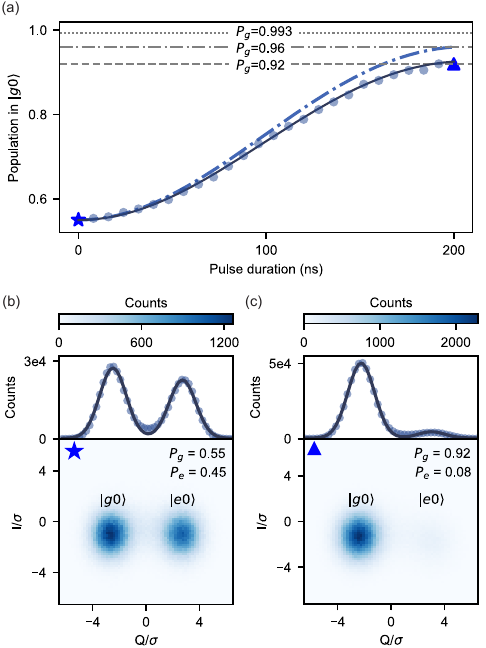}%
            \caption{\label{fig:3}  
            (a) Measured qubit $|g0\rangle$ population vs. reset pulse duration for a fixed pulse amplitude. A maximum of $92\%$ fluxonium population is reset to the ground state (dashed) before correction for qubit decay during readout. Taking the readout error into account, the maximum population reaches 96\% (dash-dotted). A reset threshold of $99.3\%$ is predicted by numerical simulation in this system upon having a longer reset pulse with no $t_\mathrm{buffer}$ and perfect readout (dotted).   
            Histograms of single-shot qubit readout at thermal equilibrium (b) and at the maximum reset efficiency (c) with $5\times 10^5$ data points collected are provided. 
        }
        \end{figure}
        
    \section{Fast Reset}
        
        \par
        Here we perform fast reset of the fluxonium in the strong coupling regime. 
        Specifically, we turn on a strong pump for a pulse duration $t=200$~ns for fast evacuation of the $|e0\rangle$ population into the $|g1\rangle$ state. 
        We then turn off the drive and wait for $t_\mathrm{buffer} = 5\ \mu$s to allow the photon to leave before readout. 
        In Fig. \ref{fig:3}(a), we show the measured fluxonium population in $|e0\rangle$ as a function of the reset pulse duration for a fixed pulse amplitude. 
        As we increase the pulse duration, the fluxonium population in $|e0\rangle$ decreases and reaches the maximum population in $|g0\rangle$ with pulse duration $t=200~$ns before the excitation gets swapped back to $|g0\rangle$. 
        At thermal equilibrium, the qubit ($\omega_{f}/2\pi=81$~MHz) has a probability of $55\%$ to be in the ground state, which is equivalent to a qubit temperature of 20 mK (Fig. \ref{fig:3}(b)). Using our reset protocol, we measure a maximum of $92\%$ qubit population in the ground state, which corresponds to cooling the qubit to 1.6 mK (Fig. \ref{fig:3}(c)). 
        Both the long timescale and the observed inefficiency of the reset is due to the low decay rate $\kappa/2\pi = 0.4$ MHz of the resonator. 
        The qubit heats during $t_\text{buffer}$ and readout. 
        If we correct for the readout error induced due to the qubit population change, the reset efficiency rises to $\sim96\%$.

        We note two possible ways to improve upon our demonstrated reset method. By prolonging the reset pump while waiting for the photon to leave the resonator, no $t_\mathrm{buffer}$ is needed and we predict the reset efficiency to saturate at $\sim 99.3\%$ for a reset pulse of duration $\gtrsim5\ \mu$s based on numerical simulations assuming perfect readout (Appendix \ref{A4}). Alternatively, a higher reset efficiency of $\sim99.8\%$ can be reached in 500~ns using a resonator with $\kappa/2\pi=4$~MHz  (Appendix \ref{A4}).
        
        \begin{figure}[t]
        \centering
        \includegraphics{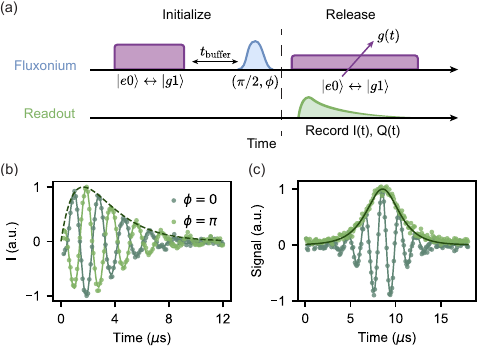}%
        \caption{\label{fig:4} 
            (a) The pulse sequence for coherently transfering an excitation from the fluxonium to a traveling photon. The fluxonium is initialized to be in $|\psi\rangle=(|g0\rangle+ e^{i\phi}|e0\rangle)/\sqrt{2}$ by first applying a reset pulse followed by a Rabi drive of amplitude $\pi/2$ and phase $\phi$ on the qubit. A photon is then released by weakly pumping the fluxonium at the $|e0\rangle\rightarrow|g1\rangle$ transition frequency such that $g \ll \kappa$. 
            (b) Measured in-phase quadrature of the emitted photon as a function of time, with the fluxonium initialized to be in
            $|\psi^\pm\rangle=(|g0\rangle\pm|e0\rangle)/\sqrt{2}$. A square pulse is used to release a photon, and the data is fit using input-output theory (solid). With a constant coupling, the output fields mostly follow an exponential decay (dashed). 
            (c) Shaping the wave packet to be a hyperbolic secant function. The measured in-phase quadrature (dark green) and field amplitude (light green) of the emitted photon is plotted as a function of time with the fluxonium initialized to be in $|\psi^+\rangle$.  
            }
        \end{figure}   
    
    \section{Controlled Release}\label{Controlled Release}
        \par 
        Equipped with the ability to initialize the fluxonium qubit state, we now demonstrate controlled transfer of a fluxonium excitation to a flying photon in the weak coupling regime. By modulating the amplitude and phase of the drive, we can shape the temporal profile of the outgoing photon wavepacket, which is useful for implementing deterministic state transfer protocols \cite{PhysRevLett.78.3221}. %

        \par %
        For characterization of the controlled conversion process, we analyze the emitted photon superposition state of the  $(|0\rangle+e^{i\phi}|1\rangle)/\sqrt{2}$, which has a non-zero field amplitude. To initialize the qubit in states $|\psi\rangle=(|g0\rangle+e^{i\phi}|e0\rangle)/\sqrt{2}$, we apply a reset pulse followed by a $\pi/2$ pulse of phase $\phi$ on the qubit. A traveling photon of state $(|0\rangle+e^{i\phi}|1\rangle)/\sqrt{2}$ is then controllably released by pumping the fluxonium at the $|e0\rangle\rightarrow|g1\rangle$ transition frequency in the regime where 
        $g \ll \kappa$ (pulse sequence shown in Fig. \ref{fig:4}(a)). The outgoing field is recorded using heterodyne detection, which also captures the phase information of the emitted photon. %
        \par
        For initial characterization, we perform this controlled release protocol using a simple square pulse. In Fig. \ref{fig:4}(b), we show the measured in-phase quadrature of the propagating fields as a function of time, with the fluxonium initialized to be in $|\psi^\pm\rangle=(|g0\rangle\pm|e0\rangle)/\sqrt{2}$. With a square pulse, the coupling rate of $|e0\rangle\leftrightarrow|g1\rangle$ is constant, which should result in a simple exponential damping of the output fields with $\Gamma_\mathrm{eff}=4g^2/\kappa$ after a ring-up time proportional to $\kappa^{-1}$ as shown in Fig. \ref{fig:4}(b). In this case, the exact response of the resonator as well as the emitted photon related by $\hat{r}_\text{out}=\sqrt{\kappa}\hat{r}$ can be analytically computed using input-output theory (Appendix \ref{A3}), which fits our measured data well. Here the coupling rate $g(t)$ is fit to be $2\pi\times 0.1\ \mathrm{MHz} \ll \kappa$. Upon shifting the phase of the qubit Rabi pulse by $\pi$, the response of the in-phase signal also shifts correspondingly, which indicates the phase coherence of the controlled release process. 

        \par
        We now demonstrate that we can dynamically tune the photon's emission rate to build a time-symmetric wave packet, which can be used in remote entanglement protocols for efficient absorption by a receiver. We shape the outgoing wave packet to a hyperbolic secant function, 
        \begin{equation}
            \label{shape}
            \Psi(t)= \frac{1}{4}\sqrt{\gamma_\text{ph}}\text{sech}(\frac{\gamma_\text{ph}t}{2}),
        \end{equation}
        where $\gamma_\text{ph}$ is the photon bandwidth. %
        In Fig. \ref{fig:4}(c), we show the measured in-phase quadrature and field amplitude of the shaped wave packet with the fluxonium initialized to be in $|\psi^+\rangle$, which displays a clear hyperbolic secant shape.
        Based on numerical simulation, we estimate a conversion efficiency of $\eta\approx79\%$ taking into account state preparation errors, which is mainly due to qubit depolarization during the release process. By using a resonator of higher decay rate $\kappa/2\pi=4 \mathrm{\ MHz}$, the conversion process can be shortened down to $\lesssim1\ \mu$s (see Fig. \ref{fig:S_efficiency} in Appendix \ref{A5}). In this case, conversion loss due to qubit depolarization can be easily suppressed with no additional system optimization and we simulate a conversion efficiency of $\eta\approx99\%$ assuming our current qubit lifetimes and perfect state initialization. We quantify the symmetry of the wavepacket by computing its normalized scalar product with the time-reversed copy of itself \cite{PhysRevX.4.041010}, which yields $s=0.975$ (Eq. (\ref{symmetry})). 
        This emitted wavepacket could then be captured by a receiver by time-reversing the emission process \cite{PhysRevLett.78.3221}. 

    \section{Outlook}
    We have demonstrated parametric conversion between a fluxonium qubit excitation and a flying photon by successfully driving the $|e0\rangle\leftrightarrow|g1\rangle$ transition. Utilizing this transition, we have initialized the fluxonium qubit in the ground state with measured probability $>95\%$ and predict a reset threshold in the same system of $>99\%$ with a longer reset pulse. By using a resonator with a higher decay rate, our simulations indicate that the fluxonium could be reset with fidelity $\sim99.8\%$ in 500~ns. 
    We have also demonstrated coherent release and temporal shaping of a flying photon detected using heterodyne measurement with total efficiency of $\sim79\%$. An improved conversion efficiency of $\sim99\%$ can be straightforwardedly obtained by increasing the resonator decay rate.
    By introducing a microwave-photonic interface for the fluxonium qubit, we have paved the way for remotely entangling fluxonium qubits via a microwave-photonic link.

    \begin{acknowledgments} 
        We acknowledge useful discussions with Xi Cao, Michael Mollenhauer, Rafael Goncalves, Sonia Rani, Cheeranjeev Purmessur, Supriya Mandal, Shyam Shankar, and Srivatsan Chakram. Our traveling-wave parametric amplifier is provided by IBM. We are grateful to Chunyang Ding and David Schuster for providing us with PCBs. This research is carried out in part in the Materials Research Lab Central Facilities and the Holonyak Micro and Nanotechnology Lab, University of Illinois. This research is partially supported by the Air Force Office of Scientific Research under award number FA9550-21-1-032. Research is also sponsored by the Army Research Office and is accomplished under Grant Number W911NF-23-1-0096. The views and conclusions contained in this document are those of the authors and should not be interpreted as representing the official policies, either expressed or implied, of the Army Research Office or the U.S. Government. 
    \end{acknowledgments}

    \appendix
    \section{Device characterization}
        \label{A1}
        
        \begin{figure}
            \centering            \includegraphics{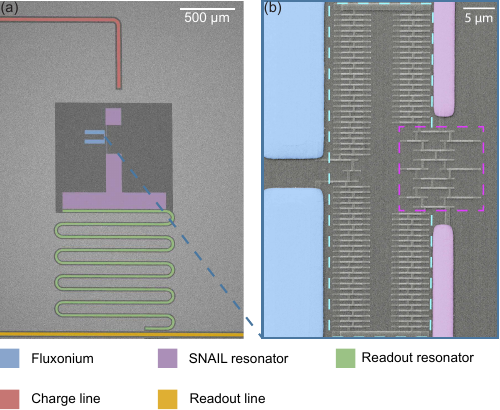}
            \caption{
             False-colored SEM images of (a) the full device and (b) the SNAIL-fluxonium. The SNAIL-fluxonium (purple and blue) couples to the readout resonator (green) via shared capacitance between the SNAIL's antenna and the readout resonator. The readout resonator couples capacitively to the readout line (orange). The charge line (red) couples capacitively to the other antenna of the SNAIL resoantor.}
            \label{fig:S_sem}
        \end{figure}
        \par
\begin{figure}[t]
            \centering
            \includegraphics{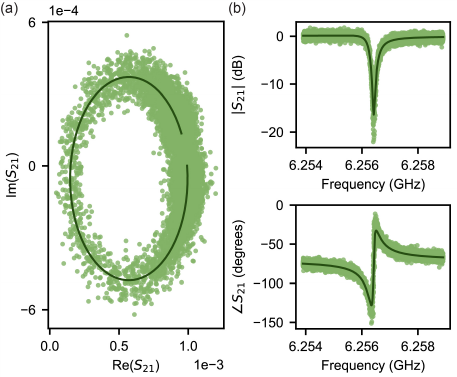}
            \caption{
            (a) $S_{21}$ response of the readout resonator at single photon level and zero flux plotted in the complex plane. (b) The same $S_{21}$ data plotted in magnitude (top) and phase (bottom).}
            \label{fig:S_21}
        \end{figure}
\begin{figure}[t]
            \centering
            \includegraphics{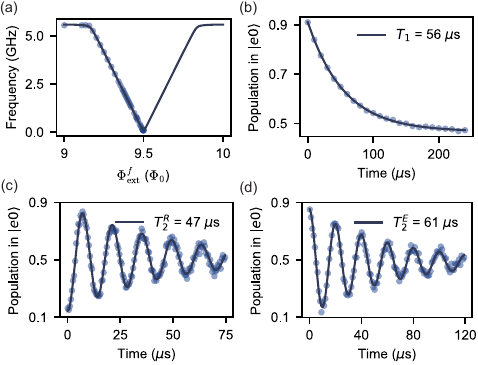}
            \caption{(a) Fluxonium qubit $|g\rangle\leftrightarrow|e\rangle$ transition frequency as a function of applied flux. Two-tone spectroscopy data is fit using the following parameters $\{E_{C,f}, E_{L,f}, E_{J,f}\}/2\pi = \{ 0.9, 0.5, 5.54\}$ GHz. 
            (b)-(d) Time domain measurements of $T_1$, $T_2^R$ and $T_2^E$ respectively.}
            \label{fig:S_time_domain}
        \end{figure}

        Figure \ref{fig:S_sem} shows the optical and scanning electron microscope (SEM) images of our device.  To drive the parity-forbidden transition \convert~, we need strong coupling between the fluxonium and the SNAIL, and between the SNAIL and the readout resonator respectively. We realize strong coupling by using multiple SNAILs as a shared inductance with the fluxonium superinductance. Moreover, one side of the SNAIL capacitor is enlarged for stronger coupling capacitance with the readout resonator and asymmetric driving of the system. A charge line weakly coupled to the other pad of the SNAIL resonator is included for driving. The coupling between the charge line and the SNAIL is designed as a trade off between the need to strongly drive the parity-forbidden transitions and while still suppressing the Purcell loss of the fluxonium excitation through the SNAIL to the charge line. The readout resonator is coupled to the readout line in hanger style with relatively low decay rate $\kappa/2\pi = 0.4$ MHz. 
        Single-tone spectroscopy for the readout resonator and two-tone spectroscopy for the fluxonium qubit is shown in Fig. \ref{fig:S_21} and Fig. \ref{fig:S_time_domain}(a) respectively.
        The resonator transmission coefficient $S_{21}(f)$ in the complex plane \cite{McRae_2020} is fit to a standard hanger geometry response:
        \begin{equation}
            S_{21}(f)=ae^{i\alpha}(1-\frac{Q/Q_c}{1+2iQ\frac{f-f_0}{f_0}}),
        \end{equation}
        where $f_0$ is the resonance frequency, $Q$ and $Q_c$ denote the total and external quality factor, $a$ attenuates the $S_{21}$ radius while $\alpha$ rotates the $S_{21}$ in the complex plane. The fluxonium spectroscopy is fit by numerically diagonalizing the Hamiltonian in Eq.~1.
        A summary of our device's parameters at $\Phi_\text{ext}^f/\Phi_0=9.5$ are provided in Table \hyperref[table:system_parameter]{I}.  
        
        We fabricate the entire device, with the exception of Al/AlOx/Al junctions, with tantalum on a sapphire wafer. A global coil is used to tune the flux across the entire chip, including flux threading though the fluxonium $\Phi_\text{ext}^f$ and the SNAIL $\Phi_\text{ext}^s$. This is made possible by designing a large area ratio between the fluxonium and the SNAIL $A_f/A_s\approx35$, so that one may tune either flux independently. We operate our device at $\Phi_\text{ext}^f/\Phi_0=9.5$, corresponding to $\Phi_\text{ext}^s/\Phi_\text{0}=0.27$.

        \begin{table}\label{table:system_parameter}
        \caption{System parameters at $\Phi_\text{ext}^f/\Phi_0=9.5$. Parameters for the fluxonium and the readout resonator are extracted by fitting their respective spectroscopy. The coupling strengths are estimated based on the measured dispersive shift. We provide designed values for parameters we do not measure.}
        \begin{threeparttable}
        \centering
        \def\arraystretch{1.2}
        \begin{tabular}{c@{\hskip 0.2in}c@{\hskip 0.2in}|c@{\hskip 0.2in}c@{\hskip 0.2in}}
        \hline\hline
        \multicolumn{2}{c|}{Fluxonium}&\multicolumn{2}{c}{Readout Resonator}\\
          $E_{C,f}/2\pi$ & 0.89 GHz & $\omega_r/2\pi$ & 6.245 GHz \\
          $E_{L,f}/2\pi$ & 0.47 GHz & $\kappa/2\pi$ & 0.4 MHz \\
          $E_{J,f}/2\pi$ & 5.54 GHz & $\chi_{fr}/2\pi$ & 0.55 MHz\\ 
          $\omega_q/2\pi$ & 81 MHz & & \\ 
        \hline\hline
        \multicolumn{2}{c|}{Single SNAIL}&\multicolumn{2}{c}{SNAIL Resonator}\\
        $\alpha$ & 0.5\tnote{*} & $E_{C,s}/2\pi$ & 0.39\tnote{*}\hspace{1.5ex}GHz\\
        $E_{J,s}/2\pi$ & 65\tnote{*}\hspace{1.5ex}GHz& $E_{L,s}/2\pi$ & 15.4\tnote{*}\hspace{1.5ex}GHz\\
        $N_s$& 3& $c_3^s/2\pi$ & 350\tnote{*}\hspace{1.5ex}MHz\\
        \hline\hline
        $g_{fs}/2\pi$ & 500 MHz &$g_{sr}/2\pi$ & 130 MHz \\
        \hline\hline
        \end{tabular}
        \begin{tablenotes}
            \item[*] These are designed parameters, as we do not perform two-tone spectroscopy on the SNAIL mode.
        \end{tablenotes}
        \end{threeparttable}
        \end{table}

        \par       
        The time domain measurements for our fluxonium qubit are shown in Fig. \ref{fig:S_time_domain}(b)-(d). A reset pulse is added for initialization before the usual $T_1$ and $T_2$ measurements. Fits of the flux-dependence of our device's relaxation time suggest that it is mainly limited by Purcell loss through the internal loss of the SNAIL resonator.    

        \section{Experimental setup}
        \label{A2}

        \begin{figure*}
            \centering
            \includegraphics{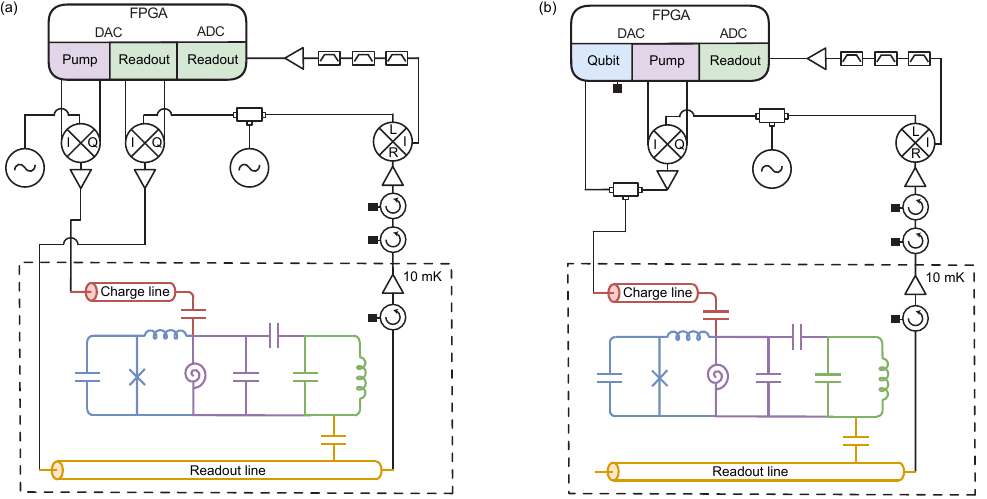}
            \caption{
            Simplified schematic of the measurement setup for (a) two-tone spectroscopy and (b) controlled release experiment. In (a), both the readout tone and the pump tone for driving the $|e0\rangle\leftrightarrow|g1\rangle$ transition are generated by mixing local oscillator (LO) from a microwave generator (\emph{SignalCore, SC5511A}) and IQ tones from an integrated FPGA system (\emph{Quantum Machines, OPX+}). The pump signal is first amplified and then sent into the dilution refrigerator (\emph{Oxford, Triton 500}) through the sample's charge line. The readout tone transmits through the readout line. The output signal from the sample goes through a circulator and is amplified by a traveling wave parametric amplifier (TWPA) at the mixing chamber (MXC). Above the MXC, the output signal goes through two additional circulators. At room temperature, the signal gets amplified before demodulated down to 50 MHz. The demodulated signal goes through three 41-58 MHz band-pass filters before a final stage of amplification. The signal is then further demodulated by analog-to-digital converters (ADC) in the FPGA system. In (b), the qubit Rabi drive is directly output by the FPGA system. The qubit Rabi drive and the pump tone are combined and sent through the charge drive line. No readout tone is applied to the system, as we are directly reading out the photon released from the readout resonator. We use the same pump LO generator for output demodulation, so that the output signal is phased locked to the output LO. }
            \label{fig:wiring}
        \end{figure*}

        \par
        Figure \ref{fig:wiring} shows the simplified schematic of the measurement setup used for the two-tone spectroscopy measurement in Section \ref{Fluxonium-SNAIL-resonator system} and the controlled-release experiment demonstrated in Section \ref{Controlled Release}.
        To perform two-tone spectroscopy, a readout tone is sent to the sample through the readout line while the amplified pump tone is sent to the qubit through the charge line weakly-coupled to the SNAIL's antenna. 
        On the output side, the transmitted readout tone is first routed through a circulator before being amplified by a TWPA. The signal goes through two additional circulators thermalized to the MXC. At room temperature, the output signal is amplified and then demodulated down to 50 MHz by the readout LO. The demodulated signal goes through 41-58 MHz band-pass filters to filter away any pump tones that leak to the readout line. After filtering, the signal gets amplified before being demodulated and digitalized by the FPGA for obtaining the I and Q components of the transmitted readout tone.  
        
        For the controlled release experiment, it is important to ensure that the photon release and photon capture processes are phase-locked. To ensure phase-locking of the release process, we apply the Rabi pulse on the qubit directly using the I channel output from the FPGA system, and modulate the pump tone addressing the $|e0\rangle\leftrightarrow|g1\rangle$ transition using the same FPGA system. To ensure phase-locking of the output signal, we use the pump signal generator to provide the LO for output demodulation. No readout tone is applied to the sample, since the photon we capture is emitted through conversion of the fluxonium excitation into a readout resonator photon.

    \section{System Hamiltonian}
        \label{A3}
        
        Without microwave drives, the Hamiltonian of the system consisting of the fluxonium qubit mode ($\hat{f}$), the SNAIL resonator mode ($\hat{s}$), and the readout resonator mode ($\hat{r}$) is given by 
        \begin{equation}
            \hat{H}/\hbar=\sum_{j=f,s,r}\omega_j\hat{j}^\dag\hat{j} + \sum_{j\neq k}\chi_{jk}\hat{j}^\dag\hat{j}\hat{k}^\dag\hat{k}\\
            + c_3^s\hat{\phi}_s^3,  
        \end{equation}
        considering only single-photon excitations of the system. Here the flux across the SNAIL is given by the sum of the contributions from all the circuit elements,  
        \begin{equation}
            \hat{\phi}_s = \sum_{j=f,s,r}\phi_{j,\mathrm{zpf}}(\hat{j}^\dag + \hat{j}), 
        \end{equation}
        where $\phi_{j,\mathrm{zpf}}$ are the quantum zero-point fluctuations of junction flux associated with each mode obtained from Hamiltonian diagonalization. For quantum systems involving the fluxonium qubit, it is difficult to obtain an analytical solution for $\chi_{kl}$ and $\phi_{j,\mathrm{zpf}}$, but one may perform numerical diagonalization to obtain these parameters. The three-wave-mixing term in the Hamiltonian breaks the parity selection rule, and allows one to drive the parity-forbidden transition $|e0\rangle\leftrightarrow|g1\rangle$.  
        
        \par
    
        In the presence of a microwave pump acting on the SNAIL, the three-wave-mixing and the driving terms are given by, 
        \begin{multline}
            \hat{H}_{\text{3WM+d}}/\hbar =c_3^s[\phi_{f,\mathrm{zpf}}(\hat{f}^\dag + \hat{f})+\phi_{s,\mathrm{zpf}}(\hat{s}^\dag + \hat{s})\\
            +\phi_{r,\mathrm{zpf}}(\hat{r}^\dag + \hat{r})]^3 + 
            2\epsilon(t)\cos{(\omega_d t)}(\hat{s}^\dag + \hat{s}),
        \end{multline}
        where $\epsilon(t)$ is the time-dependent pump amplitude. Going into the displaced frame, we introduce the displaced amplitude,
        \begin{equation}
            \tilde{\xi}(t)=\xi(t)e^{-i\omega_dt}=\frac{\epsilon(t)e^{-i\omega_dt}}{i\kappa_s/2+\Delta_c}\approx\frac{\epsilon(t)e^{-i\omega_dt}}{\Delta_c}, 
        \end{equation}
        where the detuning $\Delta_c$ is the difference between the drive frequency and the SNAIL mode frequency $\Delta_s=\omega_d-\omega_s$. Through the substitution of $\hat{s} = \hat{\tilde{s}} + \tilde{\xi}(t)$ and moving into the rotating frame, we find the conversion Hamiltonian, 
        \begin{align}
            \hat{H}_\text{conv}/\hbar &= 6c_3^s(\prod_{i=f,s,r}\phi_{i,\mathrm{zpf}})(\xi(t)\hat{f}\hat{r}^\dag e^{-i\delta t}+\xi^*(t)\hat{f}^\dag\hat{r} e^{i\delta t}) \nonumber \\
            & =(g(t) \hat{f}\hat{r}^\dag e^{-i\delta t} + g(t)^* \hat{f}^\dag\hat{r} e^{i\delta t}),
        \end{align}
        where $g(t)=6c_3^s\xi(t)\phi_{f,\mathrm{zpf}}\phi_{s,\mathrm{zpf}}\phi_{r,\mathrm{zpf}}$ is the coupling rate of the $|e0\rangle\leftrightarrow|g1\rangle$ transition, and $\delta = \omega_d-(\omega_r-\omega_f)$ is a small detuning between the pump frequency and the $|e0\rangle\leftrightarrow|g1\rangle$ transition frequency. In simulation, we find that computing the bare SNAIL phase matrix element squared  $6c_3^s\xi(t)|\langle g1|\hat{\phi}_s^2|e0\rangle|$ already gives useful guidance for choosing system parameters.

        Besides $\hat{H}_\text{conv}$, there are additional terms corresponding to the ac-Stark shifts generated by the microwave pump on the SNAIL,
        \begin{equation}
            \hat{H}_\mathrm{Stark}/\hbar=|\xi(t)|^2(\chi_{fs}\hat{f}^\dag\hat{f}+\chi_{sr}\hat{r}^\dag\hat{r}).
        \end{equation}   
        The Stark shifts can be compensated by modulating the pump frequency. Large-amplitude drives result in the Stark shift changing non-negligibly during the pulse rise time, which significantly complicates pulse calibration. We therefore choose to operate the system at high flux quanta for a large third-order coefficient $c_3^s$ and apply pumps of relatively low power. 

        In Fig.9, we show two-tone spectroscopy data of the $|e0\rangle\leftrightarrow|g1\rangle$ transition under 300 ns pump pulse as a function of pump power at the chip. In the weak coupling regime, the $|e0\rangle\leftrightarrow|g1\rangle$ transition frequency is centered at 6.16455 GHz. In the strong coupling regime, the $|e0\rangle\leftrightarrow|g1\rangle$ transition frequency is only shifted by 200 kHz due to the stronger pump tone and is centered at 6.16435 GHz. One can estimate the number of photons sent into the SNAIL via the formula \cite{Capelle_2020} 
        \begin{equation}
            |\xi|^2 = \bar{n}=2\kappa_\mathrm{ext}|a_\mathrm{in}|^2/(\kappa_\mathrm{tot}+4\Delta^2),
        \end{equation}
        where $|a_\mathrm{in}|^2$ is the incoming photon flux in unit of photon/s. Assuming internal and external quality factors of $Q_\mathrm{in}\approx5\times10^4, Q_\mathrm{ext}\approx5\times10^5$ and an expected frequency of $\omega_s/2\pi=6.93$ GHz for the SNAIL, we estimate $|\xi|^2\approx0.088$ in the weak coupling regime and $|\xi|^2\approx0.86$ in the strong coupling regime, both of which are below single photon level. 

        Summarizing all the terms, we may write out an effective system Hamiltonian, 
        \begin{multline}
            \hat{H}_\mathrm{eff}/\hbar=\tilde{\omega}_f\hat{f}^\dag\hat{f}+\tilde{\omega}_r\hat{r}^\dag\hat{r}+\chi_{fr}\hat{f}^\dag\hat{f}\hat{r}^\dag\hat{r} \\
            +2g\cos(\omega_dt)(\hat{f}\hat{r}^\dag+\hat{f}^\dag\hat{r}),
        \end{multline}
        where $\tilde{\omega_f},\tilde{\omega_r}$ is the new fluxonium and resonator frequency including Stark shifts. Here we omit the SNAIL mode in the effective Hamiltonian, since the SNAIL mode should not be populated in our experiment. We also assume $g$ to be real without loss of generality.   

        \par
        We perform more precise simulations of the time evolution of the given quantum system by solving the Lindblad master equation,
        \begin{equation}\label{master_equation}
            \delta_t\hat{\rho}=-i[\hat{H}_\text{eff}, \hat{\rho}] + \sum_{k=g0,e0,g1}[L_k\rho L_k-\frac{1}{2}\{L_kL_k,\hat{\rho}\}].
        \end{equation}
         Here $\hat{\rho}$ is the system density operator. $\hat{H}_\text{sys}$ is defined in equation (\ref{Hamiltonian}) and $\hat{H}_d$ is the driving Hamiltonian. The jump operators $L_{g0}=\sqrt{\Gamma_{\downarrow}}|e0\rangle\langle g0|, L_{e0}=\sqrt{\Gamma_{\uparrow}}|g0\rangle\langle e0|, $ and $L_{g1}=\sqrt{\kappa}|g0\rangle\langle g1|$ describe the main depolarization paths of the quantum system. 

        \par
        Considering only the conversion $|e0\rangle\leftrightarrow|g1\rangle$ at resonance, and resonator decay $\kappa$ into the readout line, the system is well described by the Heisenberg equations of motion,  
         \begin{align}
             \label{input-output-theory}
             \delta_t \hat{f} &= -ig\hat{r}, \\
             \delta_t \hat{r} &= -ig\hat{f} - \frac{\kappa}{2}\hat{r}, \\
             \hat{r}_\text{out} &= \sqrt{\kappa}\hat{r},
         \end{align}
         where $\hat{r}_\text{out}$ is the outgoing photonic mode. For the case of constant coupling rate $g$ with some initial condition $\hat{f}(0)$, one may obtain an analytical solution. The analytical solution and discussion of the resonant case are carefully laid out in the supplemental information of Ref. \cite{2017NatPh..13..882P} for a system of two cavity modes.

        \begin{figure}[t]
            \centering            \includegraphics{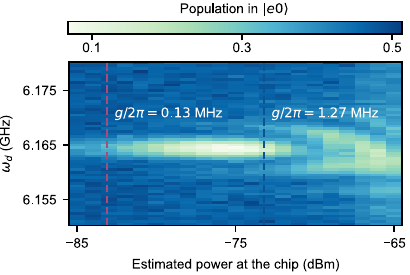}
            \caption{
            Two-tone spectroscopy data of  $|e0\rangle\leftrightarrow|g1\rangle$ transition under 300 ns pump pulse as a function of pump power estimated at the chip. Dashed lines label the different power levels we use to realize weak (red) and strong (blue) coupling demonstrated in Fig.2.}
            \label{fig:enter-label}
        \end{figure}
        
    \section{Fast reset efficiency}\label{A4}

            \begin{figure*}[t]
        \centering
        \includegraphics{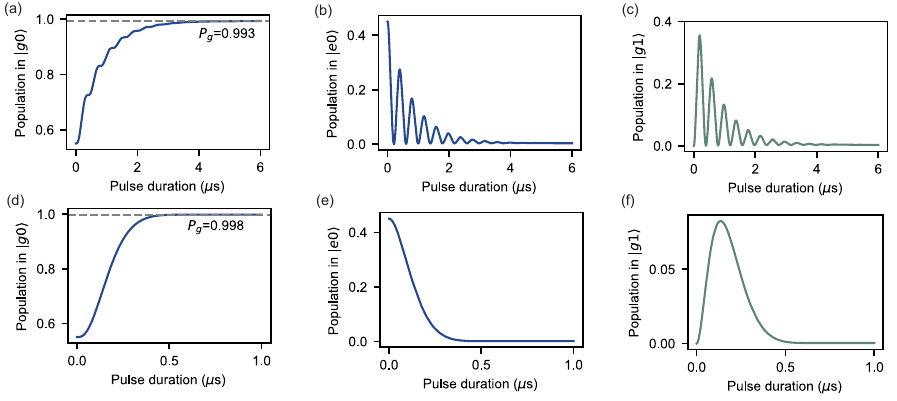}%
        \caption{\label{fig:S_cooling_with_longer_pump}
            (a)-(c) Simulated population in $|g0\rangle, |e0\rangle, |g1\rangle$ vs. reset pulse duration with the same parameters as Fig. \ref{fig:3}(a), but zero $t_\mathrm{buffer}$. A maximum of $\sim99.3\%$ fluxonium population (dashed) can be initialized to the ground state.   
            (d)-(f) Simulated population in $|g0\rangle, |e0\rangle, |g1\rangle$ vs. reset pulse duration with $g/2\pi=1.27 $ MHz and $\kappa/2\pi=4$ MHz. A higher resonator decay rate can increase the ground state reset population to $\sim99.8\%$ (dashed).
        } 
        \end{figure*}
        
        Our reset inefficiency is mainly due to qubit $T_1$ depolarization. There are three stages of depolarization (mainly heating) we need to consider for our reset process: during pumping, during $t_\mathrm{buffer}$, and during readout. Knowing the qubit depolarization rate $\Gamma$ and temperature, we can obtain $\Gamma_\uparrow$ and $\Gamma_\downarrow$ related by the following equation \cite{Krantz_2019}, 
        \begin{align}
            \Gamma &= \Gamma_\uparrow + \Gamma_\downarrow \\
            \Gamma_\uparrow &= \exp{(-\hbar\omega_f/k_BT)}\Gamma_\downarrow.
        \end{align}
        To get an estimate of the fluxonium heating during the 200 ns pumping time, we solve the master equation [Eq. (\ref{master_equation})] and find the loss to be negligible. During $t_\mathrm{buffer}$, the fluxonium population heats toward thermal equilibrium, 
        \begin{equation}\label{decay}
            P_e(t)= \Gamma_\uparrow/\Gamma + (P_{e,0}-\Gamma_\uparrow/\Gamma)\exp{(-\Gamma t)}, 
        \end{equation}
        where $P_{e,0}$ is the $|e0\rangle$ population initially after pumping. Since a heterodyne measurement integrates the signal over time, the final population we obtain from readout of pulse duration ${t}_\mathrm{readout}$ is given by, 
        \begin{equation}
            \bar{P_e} =1/{t}_\mathrm{readout}\int_{t_\mathrm{buffer}}^{t_\mathrm{buffer}+t_\mathrm{readout}}P_e(t)dt, 
        \end{equation}
        In our experiment, we use $t_\mathrm{buffer} = 5\ \mu$s and $t_\mathrm{readout} = 10\ \mu$s, which results in $\sim$ 4\% loss during $t_\mathrm{buffer}$ and another time-averaged $\sim$ 4\% loss during readout.

        One can improve the reset efficiency by leaving the pump on while waiting for $|g1\rangle$ decaying to $|g0\rangle$. In Fig. \ref{fig:S_cooling_with_longer_pump}(a)-(c), we show the simulated time evolution of the system population with a longer reset pulse, assuming perfect readout. Population in $|g0\rangle$ reaches a steady state of $\sim99.3\%$ shortly after 5 $\mu$s of pumping, denoting the highest possible reset efficiency in our current system. This reset threshold is consistent with the simple calculation, 
        \begin{align}
            P_{g}^\mathrm{th} &= \frac{\Gamma_\uparrow+\Gamma_\mathrm{eff}}{\Gamma_\uparrow + \Gamma_\downarrow +\Gamma_\mathrm{eff}}, \\
            &= \frac{\Gamma_\uparrow+\kappa/2}{\Gamma_\uparrow + \Gamma_\downarrow +\kappa/2} \approx 99.3\%,
        \end{align}
        where $\Gamma_\mathrm{eff}$ denotes the effective decay rate of the system. However, by leaving the pump on, the resonator photon decays at a rate of $\kappa/2$ instead of $\kappa$, which lengthens the initialization process. Therefore we choose to demonstrate faster reset at the expense of slightly lower ground state initialization in the main text. %
        
        If we keep the same coupling rate $g/2\pi=1.27$ MHz but use a higher decay rate for the resonator $\kappa/2\pi=4$ MHz, the system would be in the weak coupling regime. In this case, the effective decay rate is given by $4g^2/\kappa$, and no $t_\mathrm{buffer}$ is needed as the population in $|g1\rangle$ is strongly suppressed. The $|g0\rangle$ population reaches $\sim$99.8\% after 500 ns of pumping according to numerical simulation (Fig. \ref{fig:S_cooling_with_longer_pump}(d)-(f)). Despite the need for a slightly longer reset pulse, the total reset time is reduced and the reset efficiency is improved drastically.     
        
    \section{Shaping wave packets}\label{A5}

        \begin{figure}[t]
        
        \includegraphics{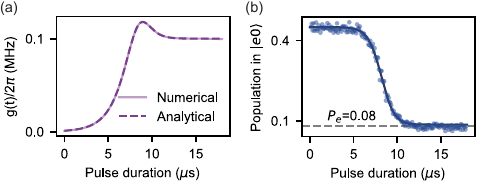}%
        \caption{\label{fig:S_release}
        (a) Coupling strength $g(t)$ as a function of pulse duration obtained from numerical simulation (solid) and analytical solution (dashed) to produce a hyperbolic secant wave packet with $\gamma_\text{ph}/2\pi=0.2$ MHz. (b) Measured fluxonium population in $|e0\rangle$ vs. pulse duration under the pump tone shown in (a). The measured population is fit using the master equation (solid). The qubit population saturates at around $8\%$, which is consistent with our reset efficiency (dashed).  
        } 
                    
         \end{figure}

        To generate our desired wave packet [Eq. (\ref{shape})], the coupling rate between the fluxonium qubit and the readout resonator can be solved analytically \cite{PhysRevA.102.053720} and should be modulated as 
        \begin{equation}
            g(t)=\frac{\gamma_\text{ph}}{4\cosh{(\frac{\gamma_\text{ph} t}{2}})}
            \frac{1-e^{\gamma_\text{ph}t} +(1+e^{\gamma_\text{ph}t})\kappa/\gamma_\text{ph}}{\sqrt{(1 + e^{\gamma_\text{ph}t} )\kappa/\gamma_\text{ph}-e^{\gamma_\text{ph}t}}},
        \end{equation}
        where the photon bandwidth bandwidth $\gamma_\text{ph}$ should be smaller than resonator decay rate $\kappa$. 
        
        \par
        While the analytical solution is specific for our chosen wave packet shape, numerical computation can be employed for different target shapes. With the equations of motion for the system, we may integrate equation (\ref{input-output-theory}), which gives
        \begin{align} 
            \hat{f}(t) &= \hat{f}(0)-i \int_{0}^{t}g(t)\hat{r}(t)dt.
        \end{align}
        Substituting it into equation \hyperref[input-output-theory]{(C9)}, we obtain 
        \begin{align}
            \delta_t \hat{r}=-ig(t)[\hat{f}(0)-i\int_{0}^{t}g(t)\hat{r}(t)dt]-\frac{\kappa}{2}\hat{r}(t).
        \end{align}
        We can rewrite the integral in terms of Riemann sum for numerical simulation purpose
        \begin{align}
         i\int_{0}^{t}g(t)\hat{r}(t)dt &\approx i\sum_{i=0}^{\frac{t}{\Delta t}-1}g(t_i)\hat{r}(t_i)\Delta t \nonumber\\
         \delta_t \hat{r} + \frac{\kappa}{2}\hat{r}(t) &\approx -ig(t)[\hat{f}(0)-i\sum_{i=0}^{\frac{t}{\Delta t}-1}g(t_i)\hat{r}(t_i)\Delta t]. 
        \end{align}
         
        Reorganizing the terms, we come to an expression for the coupling rate $g(t)$ as follows,
        \begin{align}\label{numerical}
        g(t) = -\frac{\delta_t \hat{r} + \frac{\kappa}{2}\hat{r}(t)}{i\hat{f}(0)+\sum_{i=0}^{\frac{t}{\Delta t}-1}g(t_i)\hat{r}(t_i)\Delta t}.
        \end{align}
        Given a target wave packet $\hat{r}_\text{out}(t)$, we solve for the coupling strength at any time using equation (\ref{numerical}). In the end, we can check if our simulated coupling strength allows for full evacuation of fluxonium excitation by making sure
        \begin{equation}
            \hat{f}(0)=i\int_{0}^{t}g(t)\hat{r}(t)dt.
        \end{equation}
         
        \par
        In Fig. \ref{fig:S_release}(a), we show both numerically and analytically computed coupling strength for producing a hyperbolic secant wave packet with $\gamma_\text{ph}/2\pi=0.2$ MHz. Both methods yield coupling strength $g(t)$ consistent with each other. In Fig. \ref{fig:S_release}(b), we show the time evolution of the fluxonium population in $|e0\rangle$ under our simulated pump tone without state initialization. The fluxonium population decays from thermal equilibrium and saturates at $8\%$, consistent with our reset fidelity in the strong coupling regime as well as numerical fit using the master equation. This indicates that our applied pump pulse amplitude evacuates a fluxonium excitation, the fidelity of which is limited by our device's $T_1$ decay during state initialization and controlled release processes due to our low decay rate $\kappa$. By simply designing a higher $\kappa$, one can achieve more efficient controlled release process without additional tuning of design parameters. 

        \begin{figure}[t]
             \includegraphics{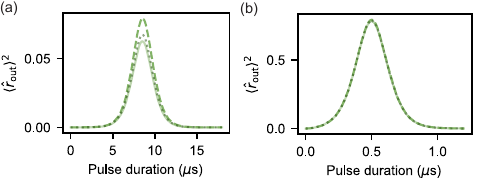}
             \caption{\label{fig:S_efficiency}
             (a) Numerically simulated field amplitude squared as a function of pulse duration for the $(|0\rangle+|1\rangle)/\sqrt{2}$ state generated in Fig. \ref{fig:4}(c). Solid line includes both state preparation error and conversion inefficiency. Dotted line corrects for state preparation error. Dashed line denotes perfect conversion efficiency. (b) Numerically simulated field amplitude squared as a function of pulse duration similar to that in Fig. \ref{fig:S_efficiency}(a) but with a higher resonator decay rate $\kappa/2\pi=4$ MHz, which allows for faster qubit-photon conversion.}
             \label{fig:enter-label}
         \end{figure}

        In Fig. \ref{fig:S_efficiency}(a), we numerically compute $\int_0^T\langle\hat{r}_\mathrm{out}(t)\rangle^2dt$ for the outgoing photonic state $(|0\rangle+|1\rangle)/\sqrt{2}$ generated in the pulse shaping experiment (Fig. \ref{fig:4}(c))  
        The conversion efficiency $\eta$ is then given by 
        \begin{equation}
            \eta =\frac{\int_0^T\langle\hat{r}_\mathrm{out}(t)\rangle^2dt}{\int_0^T\langle\hat{r}_0(t)\rangle^2dt}=4\int_0^T\langle\hat{r}_\mathrm{out}(t)\rangle^2dt,  
        \end{equation}
        where $\hat{r}_0(t)$ denotes outgoing field amplitude under perfect conversion. The total efficiency is simulated to be $\sim79\%$ including state preparation error. If correcting for state preparation error, the conversion efficiency rises to $\sim86\%$, the loss of which is mainly due to qubit depolarization during the controlled release process. However, such source of error can be easily suppressed by designing a larger resonator decay rate, which allows for a faster conversion process. With a resonator decay rate of $\kappa/2\pi=4 \mathrm {MHz}$, one can reach a conversion efficiency of $\eta\approx99\%$ under perfect state initialization (Fig. \ref{fig:S_efficiency}(b)). If we include state preparation error assuming $\sim99.8\%$ initialization efficiency (Fig. \ref{fig:S_cooling_with_longer_pump}(d)), the conversion efficiency is only compromised by an extra loss of $\sim0.4\%$.

        Finally, a flying qubit shaped into a time-symmetric profile can be absorbed with unit probability by time-reversing the release process on a second qubit $g_2(t)=g_1(-t)$ \cite{PhysRevLett.78.3221}. We quantify the symmetry of our generated wave packet by computing the following symmetry parameter \cite{PhysRevX.4.041010}
        \begin{equation}\label{symmetry}
            s=\max_{t_0}\frac{|\int\langle\hat{r}_\mathrm{out}(2t_0-t)\rangle*\langle\hat{r}_\mathrm{out}(t)\rangle dt|}{\int|\langle\hat{r}_\mathrm{out}(t)\rangle|^2dt},
        \end{equation}   
        which yields $s=0.975$. By increasing the resonator decay rate, the symmetry parameter of the shaped wave packet can be further optimized by avoiding asymmetry due to the resonator rise time and qubit depolarization. 
    \bibliography{citation}%

\nocite{*}

\end{document}